\documentclass[a4paper,11pt]{article}
\pdfoutput=1 % if your are submitting a pdflatex (i.e. if you have
\usepackage{jheppub} % for details on the use of the package, please
\usepackage[T1]{fontenc} % if needed

\def\moth{\mathsurround=0pt}
\newdimen\zo \zo=0pt

\def\tick{\leaders\hrule height 0.5ex depth 0pt \hskip 0.5pt}
\def\upboxfill{$\moth \setbox\zo\hbox{\tick}%
  \hskip 3pt\hbox to 0pt{$\tick$\hss}\hrulefill \hbox to 7.5pt{$\tick$\hss}$}

\def\dtick{\leaders\hrule height .34pt depth 0.5ex \hskip 0.5pt}
\def\downboxfill{$\moth \setbox\zo\hbox{\dtick}%
  \hskip 2pt\hbox to 0pt{$\dtick$\hss}\hrulefill \hbox to 2pt{$\dtick$\hss}$}

\def\bec{\begin{center}}
\def\ec{\end{center}}

\def\nn{\nonumber}

\def\be{\begin{equation}}
\def\ee{\end{equation}}
\def\bea{\begin{eqnarray}}
\def\eea{\end{eqnarray}}
\def\ba{\begin{array}}
\def\ea{\end{array}}

\arxivnumber{2607.24916}

\title{\boldmath Carrollian bosonic supergravity at order $\alpha'$ and the universal cancellation of higher-curvature divergences}

\author{E. Lescano}
\affiliation{Jozef Stefan Institute,\\ Jamova 39, 1000 Ljubljana, Slovenia}
\affiliation{Instituto Superior Tecnico,\\ Av. Rovisco Pais 1, 1049-001 Lisboa, Portugal}

\emailAdd{eric.lescano@ijs.si}

\abstract{We prove that the Carrollian limit of bosonic supergravity remains finite after including the complete set of four-derivative terms arising from the first $\alpha'$-correction, for an arbitrary Carrollian geometry, i.e., without imposing compatibility conditions on the Carrollian fields. The proof rests on a small set of exact identities satisfied by the leading term of the relativistic Riemann tensor, which is entirely controlled by the intrinsic torsion $K_{\mu \nu}=\frac12 \mathcal{L}_{\tau} h_{\mu \nu}$ of the Carrollian structure. We explicitly construct the finite action in a manifestly covariant form: the gravitational sector is purely electric, depending only on $K_{\mu \nu}$ and on its derivative along $\tau^{\mu}$, while the Carrollian curvature and the twist of $\tau_{\mu}$ drop out. We then show that higher-curvature contributions obey a universal scaling rule: every scalar built from $N$ relativistic Riemann tensors grows at most as $w^{2N}$ in the Carrollian expansion, so that all purely gravitational $\alpha'^{N-1}$-corrections admit a finite Carrollian limit for arbitrary Carrollian geometries. As applications we show that the $\alpha'^2$- and $\alpha'^3$-corrections of bosonic string theory, including the terms proportional to $\zeta(3)$, are finite and non-vanishing in this limit, and that the compatibility conditions $\nabla_{\mu} h_{\nu \rho}=\nabla_{\mu} \tau^{\nu}=0$ trivialize the whole gravitational sector.}

\begin{document}
\maketitle
\flushbottom

\section{Motivation}

Carrollian geometry \cite{Poincare1,Poincare2} has recently emerged as one of the central geometric frameworks in high-energy physics, including black holes \cite{BH1,BH2,BH3,BH4,BH5,BH6,BH7}, holography and effective descriptions \cite{Holo1,Holo2,Holo3,Holo4,Holo5,Holo6,Holo9,Holo10,Holo11,Holo12}. This broad interest originates from several remarkable properties. First, this geometry naturally describes the ultra-relativistic regime, where the dynamics becomes localized along a single light-cone direction while the transverse propagation is suppressed. Second, this limit considerably simplifies the underlying structure of string theory, providing a promising framework to investigate stringy effects such as higher-derivative corrections. Third, it has become a common arena where different approaches to quantum gravity can be compared within a unified language.

Here we focus on the second aspect: The low-energy limit of bosonic string theory is described by a metric, the Kalb--Ramond field and the dilaton. Integrating out the massive string modes generates an infinite tower of higher-derivative $\alpha'$-corrections built from curvature tensors (we use the Metsaev and Tseytlin formalism \cite{MetsaevTseytlin}, see \cite{Review} for a pedagogical review).

So far, it has remained unknown whether the inclusion of stringy $\alpha'$-corrections preserves the consistency of the Carrollian limit. Indeed, the higher-derivative action develops apparent divergences at intermediate stages of the expansion, making it unclear whether a finite Carrollian theory exists beyond leading order. In this work we show that the divergences cancel for an arbitrary Carrollian geometry, and that the mechanism behind the cancellation is universal: it is controlled by the intrinsic torsion of the Carrollian structure \cite{FOF}, which in the present context is the tensor $K_{\mu \nu} = \frac12 \mathcal{L}_{\tau} h_{\mu \nu}$ that also governs the leading-order (electric) Carrollian limit of general relativity \cite{BH2}.

\section{Main results}

\begin{itemize}

\item We prove that the apparent divergences of the complete four-derivative bosonic supergravity cancel exactly under the Carrollian limit for an arbitrary Carrollian geometry, extending the results of \cite{BLP}. No compatibility condition on the covariant derivatives of the Carrollian fields is required. The proof follows from a set of exact identities satisfied by the leading term $\hat R^{(2)}$ of the relativistic Riemann tensor, eqs. (\ref{idRcc1})--(\ref{idRcc4}).

\item We explicitly construct the finite Carrollian action in a manifestly covariant form, eqs. (\ref{action2}) and (\ref{action4}). All the non-metricity of the Carrollian connection is encoded in two tensors, the expansion/shear $K_{\mu \nu}$ and the twist $\omega_{\mu \nu}=\partial_{[\mu} \tau_{\nu]}$. The gravitational sector of the finite action is purely electric: it depends on $K_{\mu \nu}$ and on $\mathcal{D}_{\mu \nu} = \mathcal{L}_{\tau} K_{\mu \nu} - K_{\mu \rho} h^{\rho \sigma} K_{\sigma \nu}$ only, while the Carrollian Riemann tensor $R^{\rho}{}_{\sigma \mu \nu}(\tau,h)$ and the twist drop out.

\item We show that higher-curvature terms obey a universal scaling rule: every scalar monomial constructed from $N$ relativistic Riemann tensors, ${\rm Riem}_1 \cdots {\rm Riem}_N$, grows at most as $w^{2N}$ in the Carrollian expansion. Consequently, the purely gravitational $\alpha'^{N-1}$-corrections of string theory admit a finite Carrollian limit for arbitrary Carrollian geometries, and only the two leading terms $\hat R^{(2)}$, $\hat R^{(0)}$ of the Riemann tensor contribute to the limit.

\item As applications, the purely gravitational $\alpha'^2$- and $\alpha'^3$-corrections of bosonic string theory are finite and non-vanishing under the Carrollian limit. In particular the terms proportional to $\zeta(3)$, whose structure is shared with the type II $\alpha'^3$-corrections, survive. We also show that the compatibility conditions $\nabla_{\mu} h_{\nu \rho} = \nabla_{\mu} \tau^{\nu} =0$, equivalent to $K_{\mu \nu}=\omega_{\mu \nu}=0$, force $R^{\mu}{}_{\nu \rho \sigma} \tau^{\nu}=0$ and trivialize the gravitational sector of the Carrollian theory to all orders in $\alpha'$.
\end{itemize}

The tensor computations of this work were performed with Cadabra \cite{Cadabra1} and the resulting expressions were simplified with the assistance of Fable 5.1. All cancellations and all closed formulas presented below were additionally verified by evaluating the relativistic invariants on generic Carrollian configurations and extracting their Laurent expansion in $w$ numerically. The Cadabra code is available upon request.

\section{The setup}
The four-derivative corrections to bosonic string theory were originally derived from three- and four-point scattering amplitudes of the massless string states \cite{GS,CN}. This approach uses the string S-matrix to construct an effective Lagrangian that reproduces the corresponding low-energy interactions. The resulting Metsaev--Tseytlin effective action takes the form \cite{MetsaevTseytlin}
\bea
S = \int d^{26}x \sqrt{-\hat g} e^{-2\hat \phi} (\hat L_{\rm bos} + \frac{\alpha'}{4} \hat L_{\rm MT}) \, ,
\label{fullA}
\eea
where $\hat L_{\rm bos}$ is given by
\bea
\hat L_{\rm bos} = \hat R + 4 \partial_\mu \hat \phi \partial^{\mu} \hat \phi - \frac{1}{12} \hat H_{\mu \nu \lambda} \hat H^{\mu \nu \lambda}
\label{bos}
\eea
and
\bea
\hat L_{\rm MT} & = & \hat R_{\mu \nu \rho \sigma} \hat R^{\mu \nu \rho \sigma} - \frac12 \hat H^{\mu \nu \rho} \hat H_{\mu \sigma \lambda} \hat R_{\nu \rho}{}^{\sigma \lambda} \nn \\ && + \frac{1}{24} \hat H^4 - \frac18 \hat H^2_{\mu \nu} \hat H^{2 \mu \nu} \, ,
\label{MT}
\eea
corresponds to the first-order $\alpha'$-correction, where
\bea
\label{CSdef}
\hat H^2_{\mu \nu} & = & \hat H_{\mu}{}^{\rho \sigma} \hat H_{\nu \rho \sigma} \, , \quad
\hat H^4 =  \hat H_{\mu \nu \rho} \hat H^{\nu}{}_{\sigma \lambda} \hat H^{\sigma \delta \rho} \hat H_{\delta}{}^{\mu \lambda} \, .
\eea
Notice that the quartic invariant $\hat H^4$ is not the square of $\hat H_{\mu \nu \rho} \hat H^{\mu \nu \rho}$ but the tetrahedral contraction, in which each three-form shares exactly one index with each of the remaining three (see \cite{MN} for a review of the Metsaev--Tseytlin Lagrangian in these conventions).
Now we consider the Carrollian ansatz \cite{BLP},
\bea
\label{Metric}
\hat g_{\mu \nu} & = & h_{\mu \nu} - \frac{1}{w^2} \tau_{\mu} \tau_{\nu} \, , \\
\label{InverseMetric}
\hat g^{\mu \nu} & = & h^{\mu \nu} - w^2 \tau^{\mu} \tau^{\nu} \, , \\
\label{B-field}
\hat B_{\mu \nu} & = & b_{\mu \nu} + 2 \tau_{[\mu} A_{\nu]} \, , \\
\label{dilaton}
\hat \phi & = & \frac12 \ln{w} + \varphi \, ,
\eea
with $w=\frac{1}{c}$ and constitutive relations given by
\bea
\label{cr1}
\tau_{\mu} h^{\mu \nu} & = & \tau^{\mu} h_{\mu \nu} = 0 \, , \\
\tau_{\mu} \tau^{\mu} & = & 1 \, , \\
\tau_{\mu} \tau^{\rho} + h_{\mu \nu} h^{\nu \rho} & = & \delta_{\mu}^{\rho} \, .
\label{cr3}
\eea

The rescaling in the dilaton provides $\sqrt{-\hat g} e^{-2\hat \phi} = \frac{1}{w^2} \Omega_{c} e^{-2 \varphi}$, with $\Omega_c$ the Carrollian measure. In \cite{BLP}, it was proved that both the leading order action and the $\hat{\rm Riem}^2$ terms are finite, the latter after considering the rescaling $\alpha'= \frac{\alpha'_C}{w^2}$ which constitutes the tensionless limit of the low energy limit of string theory. Therefore, the two-derivative Lagrangian must contribute with $w^{2}$ at most, and $\hat L_{\rm MT}$ with $w^4$ at most. Since the Carrollian structure has no metric, we never raise or lower indices: all contractions are written explicitly in terms of $h^{\mu \nu}$, $h_{\mu \nu}$, $\tau^{\mu}$ and $\tau_{\mu}$. For products of symmetric spatial tensors $X_{\mu \nu}$, $Y_{\mu \nu}$ we use the shorthand $\langle X Y \rangle \equiv X_{\mu \nu} Y_{\rho \sigma} h^{\mu \rho} h^{\nu \sigma}$ and $\langle X^n \rangle \equiv X_{\mu_1 \mu_2} h^{\mu_2 \mu_3} X_{\mu_3 \mu_4} \cdots X_{\mu_{2n-1} \mu_{2n}} h^{\mu_{2n} \mu_1}$, so that $\langle X \rangle = h^{\mu \nu} X_{\mu \nu}$.

\section{Carrollian expansion of the connection and the curvature}
\label{sec:geometry}

The 3-form is decomposed as
\bea
\hat H_{\mu \nu \rho} = h_{\mu \nu \rho} + f_{\mu \nu \rho}
\eea
where the previous quantities are
\bea
h_{\mu \nu \rho} & = & 3 \partial_{[\mu} b_{\nu \rho]} \, , \\ f_{\mu \nu \rho} & = & 6 A_{[\mu} \nabla_{\nu} \tau_{\rho]} - 6 \nabla_{[\mu} A_{\nu} \tau_{\rho]} \, .
\eea
Our convention for the covariant derivative is
\bea
\nabla_{\mu} \tau_{\nu} & = & \partial_{\mu} \tau_{\nu } - \Gamma_{\mu \nu}^{\rho} \tau_{\rho} \, , \\
\nabla_{\mu} \tau^{\nu}& = & \partial_{\mu} \tau^{\nu} + \Gamma_{\mu \sigma}^{\nu} \tau^{\sigma} \, , \\
\nabla_{\mu} h_{\nu \rho} & = & \partial_{\mu} h_{\nu \rho} - 2 \Gamma_{\mu (\nu}^{\sigma} h_{\rho) \sigma} \, , \\
\nabla_{\mu} h^{\nu \rho} & = & \partial_{\mu} h^{\nu \rho} + 2 \Gamma_{\mu \sigma}^{(\nu} h^{\rho) \sigma} \, ,
\eea
where the torsionless Carrollian connection is given by
\bea
\Gamma_{\mu \nu}^{\rho} & = & \frac12 \tau^{\rho} \tau^{\sigma} \Big(2 \partial_{(\mu}(\tau_{\nu)} \tau_{\sigma}) - \partial_{\sigma}(\tau_{\mu} \tau_{\nu}) \Big) + \frac12 h^{\rho \sigma} \Big(2 \partial_{(\mu} h_{\nu) \sigma}  - \partial_{\sigma}(h_{\mu \nu}) \Big) \, .
\label{Gamma0}
\eea
The Carrollian Riemann tensor is then computed from the commutator of the covariant derivatives acting on an arbitrary vector,
\bea
[\nabla_{\mu},\nabla_{\nu}] v^{\rho} = R^{\rho}{}_{\epsilon \mu \nu} v^{\epsilon} \, , \qquad
R^{\rho}{}_{\epsilon \mu \nu}(\tau,h) = 2 \partial_{[\mu} \Gamma_{\nu] \epsilon}^{\rho} + 2 \Gamma_{[\mu| \alpha}^{\rho} \Gamma_{|\nu] \epsilon}^{\alpha} \, .
\eea

\subsection{Intrinsic torsion and the non-metricity of the Carrollian connection}
The Levi-Civita connection of the relativistic metric (\ref{Metric}) admits a finite expansion,
\bea
\hat \Gamma^{\rho}_{\mu \nu} = w^2 \Gamma^{(2)\rho}_{\mu \nu} + \Gamma^{\rho}_{\mu \nu} + w^{-2} \Gamma^{(-2)\rho}_{\mu \nu} \, ,
\label{Gammaexp}
\eea
where the $w^0$ term is the Carrollian connection (\ref{Gamma0}). Since $\Gamma^{(\pm 2)}$ are differences of connections they are tensors, and they can be written covariantly as
\bea
\label{G2}
\Gamma^{(2)\rho}_{\mu \nu} & = & - \frac12 \tau^{\rho} \tau^{\sigma} \big(\nabla_{\mu} h_{\nu \sigma} + \nabla_{\nu} h_{\mu \sigma} - \nabla_{\sigma} h_{\mu \nu} \big) = \tau^{\rho} K_{\mu \nu} \, , \\
\Gamma^{(-2)\rho}_{\mu \nu} & = & - \frac12 h^{\rho \sigma} \big(\tau_{\nu} \nabla_{\mu} \tau_{\sigma} + \tau_{\mu} \nabla_{\nu} \tau_{\sigma} - \tau_{\mu} \nabla_{\sigma} \tau_{\nu} - \tau_{\nu} \nabla_{\sigma} \tau_{\mu} \big) \, ,
\label{Gm2}
\eea
where we have introduced the extrinsic curvature of the Carrollian structure,
\bea
K_{\mu \nu} \equiv \frac12 \mathcal{L}_{\tau} h_{\mu \nu} = \frac12 \Big(\tau^{\sigma} \nabla_{\sigma} h_{\mu \nu} + h_{\sigma \nu} \nabla_{\mu} \tau^{\sigma} + h_{\mu \sigma} \nabla_{\nu} \tau^{\sigma} \Big) \, , \qquad K_{\mu \nu} \tau^{\nu} = 0 \, .
\label{Kdef}
\eea
The tensor $K_{\mu \nu}$ is the intrinsic torsion of the Carrollian structure $(\tau^{\mu}, h_{\mu \nu})$ \cite{FOF,Hartong}: a torsion-free connection compatible with the Carrollian structure exists if and only if $K_{\mu \nu}=0$. It is also the object that controls the leading-order (electric) Carrollian limit of general relativity \cite{BH2}.

The compatibility of the relativistic metric with the relativistic connection, $\hat \nabla \hat g_{\mu \nu} = \hat \nabla \hat g^{\mu \nu}=0$, implies the following relations for the covariant derivatives of the Carrollian fields,
\footnotesize
\bea
\label{metricity1}
\nabla_{\mu} h_{\nu \rho} & = & \tau^{\sigma}(\nabla_{\mu}  h_{(\nu| \sigma} + \nabla_{(\nu|} h_{\mu \sigma} - \nabla_{\sigma} h_{\mu (\nu})\tau_{\rho)} \\
\nabla_{\mu} h^{\nu \rho} & = & - 2 h^{(\nu|\sigma} (\tau_{(\mu} \nabla_{\epsilon)} \tau_{\sigma} - \tau_{(\mu} \nabla_{\sigma)} \tau_{\epsilon)} \tau^{\rho)} \tau^{\epsilon}\, ,
\\
\nabla_{\mu} (\tau^{\nu} \tau^{\rho}) & = & - \tau^{\sigma} \tau^{(\nu}(\nabla_{\mu} h_{\epsilon \sigma} + \nabla_{\epsilon} h_{\mu \sigma} - \nabla_{\sigma}h_{\mu \epsilon})h^{\rho) \epsilon} \, , \\
\nabla_{\mu} (\tau_{\nu} \tau_{\rho}) & = & h^{\epsilon \sigma} (\tau_{\mu} \nabla_{(\nu} \tau_{\sigma} + \tau_{(\nu} \nabla_{\mu} \tau_{\sigma} \nn \\ && - \tau_{\mu} \nabla_{\sigma} \tau_{(\nu} - \tau_{(\nu} \nabla_{\sigma} \tau_{\mu}) h_{\epsilon \rho)} \, .
\label{metricity4}
\eea
\normalsize
These relations can be solved explicitly. Together with the constitutive relations (\ref{cr1})-(\ref{cr3}) they fix the whole non-metricity of the Carrollian connection in terms of $K_{\mu \nu}$ and of the twist $\omega_{\mu \nu} \equiv \partial_{[\mu} \tau_{\nu]} = \nabla_{[\mu} \tau_{\nu]}$,
\bea
\label{rule1}
\nabla_{\mu} \tau^{\nu} & = & K_{\mu \sigma} h^{\sigma \nu} \, , \\
\label{rule2}
\nabla_{\mu} h_{\nu \rho} & = & - K_{\mu \nu} \tau_{\rho} - K_{\mu \rho} \tau_{\nu} \, , \\
\label{rule3}
\nabla_{\mu} \tau_{\nu} & = & \omega_{\mu \nu} - \tau_{\mu} \omega_{\nu \sigma} \tau^{\sigma} - \tau_{\nu} \omega_{\mu \sigma} \tau^{\sigma} \, , \\
\label{rule4}
\nabla_{\mu} h^{\nu \rho} & = & - 2 h^{\sigma (\nu} \tau^{\rho)} \Big(\omega_{\mu \sigma} + \frac12 \tau_{\mu} \tau^{\epsilon} \omega_{\epsilon \sigma} \Big) \, .
\eea
In particular $\tau^{\mu} \nabla_{\mu} \tau^{\nu}=0$, $\tau_{\nu} \nabla_{\mu} \tau^{\nu} = \tau^{\nu} \nabla_{\mu} \tau_{\nu} = 0$. The derivatives of the Carrollian fields are therefore not independent data: $\nabla_{\mu} \tau^{\nu}$ and $\nabla_{\mu} h_{\nu \rho}$ are completely determined by $K_{\mu \nu}$, while $\nabla_{\mu} \tau_{\nu}$ and $\nabla_{\mu} h^{\nu \rho}$ are determined by $\omega_{\mu \nu}$. The compatibility conditions $\nabla_{\mu} h_{\nu \rho}=0$, $\nabla_{\mu} \tau^{\nu}=0$ are thus equivalent to $K_{\mu \nu}=0$, and we will see in section \ref{sec:universal} that they trivialize the gravitational sector of the Carrollian theory. Throughout this work no such condition is imposed: $K_{\mu \nu}$ and $\omega_{\mu \nu}$ are arbitrary.

\subsection{Expansion of the Riemann tensor}
Writing $\hat \Gamma = \Gamma + S$ with $S = w^2 \Gamma^{(2)} + w^{-2} \Gamma^{(-2)}$ a tensor, the relativistic Riemann tensor is exactly
\bea
\hat R^{\rho}{}_{\epsilon \mu \nu} = R^{\rho}{}_{\epsilon \mu \nu}(\tau,h) + \nabla_{\mu} S^{\rho}_{\nu \epsilon} - \nabla_{\nu} S^{\rho}_{\mu \epsilon} + S^{\rho}_{\mu \alpha} S^{\alpha}_{\nu \epsilon} - S^{\rho}_{\nu \alpha} S^{\alpha}_{\mu \epsilon} \, ,
\eea
and, since $\Gamma^{(2)\rho}_{[\mu|\alpha} \Gamma^{(2)\alpha}_{|\nu]\epsilon} = \tau^{\rho} K_{[\mu| \alpha} \tau^{\alpha} K_{|\nu] \epsilon}=0$, the expansion stops at order $w^2$,
\bea
\hat R^{\rho}{}_{\epsilon \mu \nu} & = & w^{2} \hat R^{(2)\rho}{}_{\epsilon \mu \nu} + \hat R^{(0)\rho}{}_{\epsilon \mu \nu} + w^{-2} \hat R^{(-2)\rho}{}_{\epsilon \mu \nu} + w^{-4} \hat R^{(-4)\rho}{}_{\epsilon \mu \nu} \, ,
\label{Riemannfull}
\eea
with
\bea
\label{Rcc}
\hat R^{(2)\rho}{}_{\epsilon \mu \nu} & = & 2 \nabla_{[\mu} \big(\tau^{\rho} K_{\nu] \epsilon} \big) = 2 K_{[\mu| \sigma} h^{\sigma \rho} K_{|\nu] \epsilon} + 2 \tau^{\rho} \nabla_{[\mu} K_{\nu] \epsilon} \, , \\
\label{R0}
\hat R^{(0)\rho}{}_{\epsilon \mu \nu} & = & R^{\rho}{}_{\epsilon \mu \nu}(\tau,h) + 2 \Gamma^{(2)\rho}_{[\mu|\alpha} \Gamma^{(-2)\alpha}_{|\nu]\epsilon} + 2 \Gamma^{(-2)\rho}_{[\mu|\alpha} \Gamma^{(2)\alpha}_{|\nu]\epsilon} \, , \\
\hat R^{(-2)\rho}{}_{\epsilon \mu \nu} & = & 2 \nabla_{[\mu} \Gamma^{(-2)\rho}_{\nu] \epsilon} \, , \qquad
\hat R^{(-4)\rho}{}_{\epsilon \mu \nu} = 2 \Gamma^{(-2)\rho}_{[\mu|\alpha} \Gamma^{(-2)\alpha}_{|\nu]\epsilon} \, .
\eea
The leading term (\ref{Rcc}) is the only source of divergences and it is controlled exclusively by $K_{\mu \nu}$: it vanishes if and only if the intrinsic torsion vanishes. Note also that the $w^0$ term (\ref{R0}) differs from the Carrollian Riemann tensor by cross terms bilinear in $K_{\mu \nu}$ and $\omega_{\mu \nu}$, which vanish when $K_{\mu \nu}=0$.

The central technical result of this work is that $\hat R^{(2)}$ satisfies the following exact identities, valid for an arbitrary Carrollian geometry,
\bea
\label{idRcc1}
\hat R^{(2)\rho}{}_{\epsilon \mu \nu} \tau^{\epsilon} & = & 0 \, , \\
\label{idRcc2}
\hat R^{(2)\rho}{}_{\epsilon \mu \nu} \tau^{\mu} & = & \tau^{\rho} \mathcal{D}_{\nu \epsilon} \, , \\
\label{idRcc3}
h_{\rho \alpha} \hat R^{(2)\rho}{}_{\epsilon \mu \nu} & = & K_{\mu \alpha} K_{\nu \epsilon} - K_{\nu \alpha} K_{\mu \epsilon} \, , \\
\label{idRcc4}
\tau_{\rho} \hat R^{(2)\rho}{}_{\epsilon \mu \nu} & = & \nabla_{\mu} K_{\nu \epsilon} - \nabla_{\nu} K_{\mu \epsilon} \, ,
\eea
where we have defined the symmetric and spatial tensor
\bea
\mathcal{D}_{\mu \nu} \equiv \tau^{\rho} \nabla_{\rho} K_{\mu \nu} + K_{\mu \rho} h^{\rho \sigma} K_{\sigma \nu} = \mathcal{L}_{\tau} K_{\mu \nu} - K_{\mu \rho} h^{\rho \sigma} K_{\sigma \nu} \, , \qquad \mathcal{D}_{\mu \nu} \tau^{\nu} = 0 \, .
\label{Ddef}
\eea
The identity (\ref{idRcc1}) follows from (\ref{Kdef}) and (\ref{rule1}), since $\tau^{\epsilon} \nabla_{\mu} K_{\nu \epsilon} = - K_{\nu \epsilon} K_{\mu \sigma} h^{\sigma \epsilon}$ is symmetric in $\mu \nu$; the identity (\ref{idRcc2}) uses in addition $\tau^{\mu} \nabla_{\mu} \tau^{\rho}=0$, while (\ref{idRcc3}) and (\ref{idRcc4}) are immediate from (\ref{Rcc}). In words: a $\tau^{\epsilon}$ on the first lower index of $\hat R^{(2)}$ kills it, a $\tau^{\mu}$ on the antisymmetric pair projects its upper index onto $\tau^{\rho}$, and the spatial part of its upper index is quadratic in $K$. For the subleading term we will also need
\bea
\label{idR0}
h_{\rho \alpha} \hat R^{(0)\rho}{}_{\epsilon \mu \nu} \tau^{\epsilon} = h_{\rho \alpha} R^{\rho}{}_{\epsilon \mu \nu} \tau^{\epsilon} = 2 h_{\rho \alpha} \nabla_{[\mu} \big( K_{\nu] \sigma} h^{\sigma \rho} \big) \, , \qquad h_{\rho \alpha} \hat R^{(0)\rho}{}_{\epsilon \mu \nu} \tau^{\epsilon} \tau^{\mu} = \mathcal{D}_{\nu \alpha} \, ,
\eea
where the first equality states that the cross terms in (\ref{R0}) are proportional to $\tau^{\rho}$ once contracted with $\tau^{\epsilon}$, and the second uses $R^{\rho}{}_{\epsilon \mu \nu} \tau^{\epsilon} = [\nabla_{\mu},\nabla_{\nu}] \tau^{\rho}$ together with (\ref{rule1}).

\section{Finiteness of the full four-derivative bosonic action}
\label{sec:finiteness}
We now extract all the potentially divergent terms of the Metsaev--Tseytlin action. Since the three-form does not carry powers of $w$ and the two-form contributions $\hat H^4$, $\hat H^2_{\mu \nu} \hat H^{2 \mu \nu}$ contain at most two $\tau \tau$ contractions (a third one would act on an antisymmetric pair), the divergences can only come from $\hat{\rm Riem}^2$ and $\hat{\rm Riem} \hat H \hat H$.

\paragraph{$\hat{\rm Riem}^2$.} Expanding $\hat R_{\mu \nu \rho \sigma} \hat R^{\mu \nu \rho \sigma}$ with (\ref{Riemannfull}) and (\ref{Metric})-(\ref{InverseMetric}), the terms of order $w^6$ and $w^8$ are
\bea
\big[\hat{\rm Riem}^2\big]^{(6)} & = & - 2 \hat R^{(2)\mu}{}_{\nu \rho \sigma} \hat R^{(2)\gamma}{}_{\epsilon \delta \lambda} \tau^{\rho} \tau^{\delta} h_{\mu \gamma} h^{\nu \epsilon} h^{\sigma \lambda}
- \hat R^{(2)\mu}{}_{\nu \rho \sigma} \hat R^{(2)\gamma}{}_{\epsilon \delta \lambda} \tau^{\nu} \tau^{\epsilon} h_{\mu \gamma} h^{\rho \delta} h^{\sigma \lambda} \nn \\ &&
- 2 \hat R^{(2)\mu}{}_{\nu \rho \sigma} \hat R^{(2)\gamma}{}_{\epsilon \delta \lambda} \tau_{\mu} \tau_{\gamma} \tau^{\nu} \tau^{\rho} \tau^{\epsilon} \tau^{\delta} h^{\sigma \lambda}
- 4 \hat R^{(0)\mu}{}_{\nu \rho \sigma} \hat R^{(2)\gamma}{}_{\epsilon \delta \lambda} \tau^{\nu} \tau^{\rho} \tau^{\epsilon} \tau^{\delta} h_{\mu \gamma} h^{\sigma \lambda} \, , \nn \\
\big[\hat{\rm Riem}^2\big]^{(8)} & = & 2 \hat R^{(2)\mu}{}_{\nu \rho \sigma} \hat R^{(2)\gamma}{}_{\epsilon \delta \lambda} \tau^{\nu} \tau^{\rho} \tau^{\epsilon} \tau^{\delta} h_{\mu \gamma} h^{\sigma \lambda} \, .
\label{Riem2div}
\eea
Every term contains either $\hat R^{(2)\mu}{}_{\nu \rho \sigma} \tau^{\nu}$, which vanishes by (\ref{idRcc1}), or $h_{\mu \gamma} \hat R^{(2)\mu}{}_{\nu \rho \sigma} \tau^{\rho}$, which vanishes by (\ref{idRcc2}) and $h_{\mu \gamma} \tau^{\mu}=0$. Hence $\hat{\rm Riem}^2$ is finite for an arbitrary Carrollian geometry, recovering the result of \cite{BLP} without any computation.

\paragraph{$\hat{\rm Riem} \hat H \hat H$.} The only divergent contributions are of order $w^6$,
\bea
\Big[- \frac12 \hat H^{\mu \nu \rho} \hat H_{\mu \sigma \lambda} \hat R_{\nu \rho}{}^{\sigma \lambda} \Big]^{(6)} & = & - \frac12 \hat H_{\mu \nu \rho} \hat H_{\sigma \gamma \epsilon} \hat R^{(2)\mu}{}_{\delta \lambda \alpha} \tau^{\nu} \tau^{\sigma} \tau^{\delta} \tau^{\alpha} h^{\rho \gamma} h^{\epsilon \lambda} \nn \\ && + \frac12 \hat H_{\mu \nu \rho} \hat H_{\sigma \gamma \epsilon} \hat R^{(2)\mu}{}_{\delta \lambda \alpha} \tau^{\nu} \tau^{\sigma} \tau^{\delta} \tau^{\lambda} h^{\rho \gamma} h^{\epsilon \alpha} \, ,
\label{divergence}
\eea
and both terms vanish identically by (\ref{idRcc1}), since $\tau^{\delta}$ acts on the first lower index of $\hat R^{(2)}$.

We conclude that the complete four-derivative action is finite under the Carrollian limit for an arbitrary Carrollian geometry, without imposing any condition on the non-metricities (\ref{rule1})-(\ref{rule4}). We stress that this cancellation is not a consequence of power counting: the naive counting allows terms of order $w^{8}$ in $\hat{\rm Riem}^2$ and $w^{6}$ in $\hat{\rm Riem} \hat H \hat H$, and their absence is entirely due to the structure (\ref{Rcc}) of $\hat R^{(2)}$. As an independent check, we have evaluated the four invariants of (\ref{MT}) on generic Carrollian configurations $(\tau_{\mu}, h_{\mu \nu}, \hat H_{\mu \nu \rho})$ with $K_{\mu \nu} \neq 0$, $\omega_{\mu \nu} \neq 0$ in several dimensions and extracted their Laurent expansion in $w^2$ numerically: the coefficients of $w^6$ and $w^8$ vanish to machine precision, while the coefficient of $w^4$ agrees with the closed formulas of the next section.

\section{Explicit action in covariant form}
\label{sec:action}
Using (\ref{rule1})-(\ref{rule4}) and the identities (\ref{idRcc1})-(\ref{idR0}) the finite action can be written in a compact covariant form, without expanding $\hat R^{(2)}$ and $\hat R^{(0)}$. We denote $\hat H_{\tau \mu \nu} \equiv \tau^{\rho} \hat H_{\rho \mu \nu}$.

\paragraph{Two-derivative action.} The order $w^2$ part of $\hat R = \hat g^{\epsilon \nu} \hat R^{\mu}{}_{\epsilon \mu \nu}$ is $h^{\epsilon \nu} \hat R^{(2)\mu}{}_{\epsilon \mu \nu} - \tau^{\epsilon} \tau^{\nu} \hat R^{(0)\mu}{}_{\epsilon \mu \nu}$, and using (\ref{Rcc}) and (\ref{idR0}) one finds $\big[\hat R\big]^{(2)} = \langle K^2 \rangle - \langle K \rangle^2 + 2 \nabla_{\mu} (\tau^{\mu} \langle K \rangle)$. The leading order action, after taking the limit $w \rightarrow \infty$, is
\bea
S & = & \int d^{26}x \, \Omega_{c} \, e^{-2\varphi} \Big(\langle K^2 \rangle - \langle K \rangle^2 + 2 \nabla_{\mu}(\tau^{\mu} \langle K \rangle) - 4 \tau^{\mu} \tau^{\nu} \partial_{\mu} \varphi \partial_{\nu} \varphi
\nn \\ && \qquad \qquad \qquad \quad + \frac{1}{4} \hat H_{\tau \mu \nu} \hat H_{\tau \rho \sigma} h^{\mu \rho} h^{\nu \sigma} + \frac{\alpha'_{c}}{4} L_{\textrm MT}^{(4)} \Big)\, ,
\label{action2}
\eea
where the $\hat H$-term reproduces the three $h h$, $h f$, $f f$ contributions of \cite{BLP} once $\hat H_{\mu \nu \rho} = (h + f)_{\mu \nu \rho}$ is inserted. The gravitational sector is the electric Carrollian gravity Lagrangian of \cite{BH2}, up to the term $2 \nabla_{\mu}(\tau^{\mu} \langle K \rangle)$ which, since $\Gamma^{\mu}_{\mu \nu} = \partial_{\nu} \ln \Omega_{c}$, can be integrated by parts into the dilaton coupling $4 \langle K \rangle \tau^{\mu} \partial_{\mu} \varphi$. When $K_{\mu \nu}=0$ the gravitational sector reduces to $- \tau^{\epsilon} \tau^{\nu} R_{\epsilon \nu}$, with $R_{\sigma \nu} = R^{\mu}{}_{\sigma \mu \nu}$, and the Carrollian curvature only appears through this term.

\paragraph{Four-derivative action.} The finite part of $\hat L_{\rm MT}$ is
\bea
L_{\rm MT}^{(4)} & = & [\hat R_{\mu \nu \rho \sigma} \hat R^{\mu \nu \rho \sigma}]^{(4)} - \frac12 [\hat H^{\mu \nu \rho} \hat H_{\mu \sigma \lambda} \hat R_{\nu \rho}{}^{\sigma \lambda}]^{(4)} + \frac{1}{24} [\hat H^4]^{(4)} - \frac18 [\hat H^2_{\mu \nu} \hat H^{2 \mu \nu}]^{(4)} \, ,
\label{action4}
\eea
with
\bea
\label{Riem2fin}
\big[ \hat R_{\mu \nu \rho \sigma} \hat R^{\mu \nu \rho \sigma} \big]^{(4)} & = & 4 \, \langle \mathcal{D}^2 \rangle + 2 \langle K^2 \rangle^2 - 2 \langle K^4 \rangle \, , \\
\label{RHH}
 - \frac12 [\hat H^{\mu \nu \rho} \hat H_{\mu \sigma \lambda} \hat R_{\nu \rho}{}^{\sigma \lambda}]^{(4)} & = & 2 \, \hat H_{\tau \mu \nu} \hat H_{\tau \rho \sigma} \mathcal{D}_{\gamma \epsilon} h^{\mu \rho} h^{\nu \gamma} h^{\sigma \epsilon} \nn \\ && + \hat H_{\tau \mu \nu} \hat H_{\tau \rho \sigma} K_{\gamma \epsilon} K_{\delta \lambda} h^{\mu \gamma} h^{\rho \epsilon} h^{\nu \delta} h^{\sigma \lambda} \, , \\
\label{H4}
\frac{1}{24} [\hat H^4]^{(4)} & = & \frac{1}{8} \tau^{\mu} \tau^{\nu} \tau^{\rho} \tau^{\sigma} h^{\gamma \epsilon} h^{\delta \lambda} h^{\alpha \beta} h^{\chi \pi} \hat H_{\mu \gamma \alpha} \hat H_{\nu \delta \chi} \hat H_{\rho \epsilon \lambda} \hat H_{\sigma \beta \pi} \, , \\
- \frac18 [\hat H^2_{\mu \nu} \hat H^{2 \mu \nu}]^{(4)} & = &  - \frac{1}{8} \tau^{\mu} \tau^{\nu} \tau^{\rho} \tau^{\sigma} h^{\gamma \epsilon} h^{\delta \lambda} h^{\alpha \beta} h^{\chi \pi} \hat H_{\mu \gamma \delta} \hat H_{\nu \epsilon \lambda} \hat H_{\rho \alpha \chi} \hat H_{\sigma \beta \pi} \nn \\ &&- \frac{1}{2} \tau^{\mu} \tau^{\nu} \tau^{\rho} \tau^{\sigma} h^{\gamma \epsilon} h^{\delta \lambda} h^{\alpha \beta} h^{\chi \pi} \hat H_{\mu \gamma \delta} \hat H_{\nu \epsilon \alpha} \hat H_{\rho \lambda \chi} \hat H_{\sigma \beta \pi} \, .
\label{H2H2}
\eea
The expansion of (\ref{H4}) and (\ref{H2H2}) in terms of $h_{\mu \nu \rho}$ and $f_{\mu \nu \rho}$ is given in appendix A. A few comments are in order:
\begin{itemize}
\item The derivation of (\ref{Riem2fin}) proceeds as follows. The order $w^4$ part of $\hat{\rm Riem}^2$ contains eight structures, built from $\hat R^{(2)} \hat R^{(2)}$, $\hat R^{(0)} \hat R^{(2)}$, $\hat R^{(0)} \hat R^{(0)}$ and $\hat R^{(2)} \hat R^{(-2)}$. Five of them vanish by (\ref{idRcc1})-(\ref{idRcc2}), and the remaining three reduce, using (\ref{idRcc2})-(\ref{idRcc3}) and (\ref{idR0}), to $2 \langle \mathcal{D}^2 \rangle$ from $\hat R^{(2)} \hat R^{(2)} \tau_{\mu} \tau_{\gamma} \tau^{\rho} \tau^{\delta}$, to $2 \langle K^2 \rangle^2 - 2 \langle K^4 \rangle$ from $\hat R^{(2)} \hat R^{(2)}$ with all indices contracted with $h$, and to $2 R^{\mu}{}_{\nu \rho \sigma} R^{\gamma}{}_{\epsilon \delta \lambda} \tau^{\nu} \tau^{\rho} \tau^{\epsilon} \tau^{\delta} h_{\mu \gamma} h^{\sigma \lambda} = 2 \langle \mathcal{D}^2 \rangle$ from $\hat R^{(0)} \hat R^{(0)}$. Analogously, in (\ref{RHH}) the two $\hat R^{(2)}$ contributions with a single $\tau \tau$ contraction and the two $\hat R^{(0)}$ contributions with two $\tau \tau$ contractions give the same structure, $\frac12 \hat H_{\tau \mu \nu} \hat H_{\tau \rho \sigma} \mathcal{D}_{\gamma \epsilon} h^{\mu \rho} h^{\nu \gamma} h^{\sigma \epsilon}$ each, while the $\hat R^{(2)}$ term with the $\tau \tau$ contraction between the two three-forms gives the $K K$ term through (\ref{idRcc3}).
\item The finite gravitational action is purely electric: it depends on the intrinsic torsion $K_{\mu \nu}$ and on its derivative along $\tau^{\mu}$ through $\mathcal{D}_{\mu \nu}$, while the Carrollian Riemann tensor $R^{\rho}{}_{\sigma \mu \nu}(\tau,h)$ and the twist $\omega_{\mu \nu}$ cancel out. In particular $\hat R^{(-2)}$ and $\hat R^{(-4)}$ never contribute.
\item When the compatibility conditions $K_{\mu \nu}=0$ are imposed, $\hat R^{(2)}=0$, $\mathcal{D}_{\mu \nu}=0$, and the entire gravitational sector of (\ref{action2})-(\ref{action4}) vanishes, including the term $\tau^{\epsilon} \tau^{\nu} R_{\epsilon \nu}$ of the two-derivative action and the $\hat{\rm Riem} \hat H \hat H$ contribution: the only surviving four-derivative terms are (\ref{H4}) and (\ref{H2H2}). This is the content of the identity $R^{\mu}{}_{\nu \rho \sigma} \tau^{\nu} = [\nabla_{\rho},\nabla_{\sigma}] \tau^{\mu}=0$, valid for a covariantly constant $\tau^{\mu}$, which makes all the curvature contractions with $\tau^{\mu}$ on the first lower index vanish. The general case $K_{\mu \nu} \neq 0$ is therefore the physically relevant one.
\end{itemize}

\section{Universal finiteness of higher-curvature corrections}
\label{sec:universal}

The finiteness of the $\hat{\rm Riem}^{2}$ contribution is not an isolated property of the four-derivative theory. In this section we show that it extends to every purely gravitational invariant, for an arbitrary Carrollian geometry, and we identify the physical origin of this behaviour: the structure of the leading term $\hat R^{(2)}$ imposed by the intrinsic torsion.

\paragraph{Universal scaling rule.} Consider any scalar $\mathcal{I}_N$ built from $N$ relativistic Riemann tensors $\hat R^{\rho}{}_{\epsilon \mu \nu}$ by contracting indices with $\hat g_{\mu \nu}$, $\hat g^{\mu \nu}$ and Kronecker deltas. For an arbitrary Carrollian structure $(\tau_{\mu}, h_{\mu \nu})$ we find
\bea
\mathcal{I}_N = \mathcal{O}(w^{2N}) \, ,
\eea
and the coefficient of $w^{2N}$ only involves $\hat R^{(2)}$ and $\hat R^{(0)}$. Consequently, since
\bea
\sqrt{-\hat g}\,
e^{-2\hat\phi}
=
w^{-2} \Omega_{c} e^{-2 \varphi}, \quad
\alpha'^{N-1}
=
\frac{\alpha_C'^{N-1}}{w^{2N-2}},
\eea
every purely gravitational contribution $\alpha'^{N-1} \sqrt{-\hat g} e^{-2 \hat \phi} \mathcal{I}_N$ to the string effective action admits a finite Carrollian limit.

\paragraph{Origin of the scaling.} The rule follows from counting how many temporal directions can be saturated in a given contraction. Decompose every index along the Carrollian frame $(\tau^{\mu}, h^{\mu \nu})$, i.e., project each lower index either onto $\tau^{\mu}$ (a ``temporal'' index) or onto the spatial directions, and similarly for upper indices with $\tau_{\mu}$ and $h_{\mu \nu}$. By (\ref{InverseMetric}), a contraction of two lower indices with $\hat g^{\mu \nu}$ produces a factor $w^2$ if and only if both indices are temporal; by (\ref{Metric}), a contraction of two upper indices with $\hat g_{\mu \nu}$ produces $w^{-2}$ if both are temporal and $w^0$ otherwise; a Kronecker contraction never produces powers of $w$. Let $n^{\tau \tau}$ be the number of inverse metrics $\hat g^{\mu \nu}$ that contribute through their term $-w^2 \tau^{\mu} \tau^{\nu}$, i.e., that contract two temporal lower indices, and $n_{\tau \tau}$ the number of metrics $\hat g_{\mu \nu}$ that contribute through their term $-w^{-2} \tau_{\mu} \tau_{\nu}$, i.e., that contract two temporal upper indices. The position of the indices on $n$ thus mirrors that of the corresponding Carrollian fields. Let also $n_{(2)}$ the number of $\hat R^{(2)}$ factors and $n_{\rm sub}=N-n_{(2)}$ the number of subleading factors $\hat R^{(k)}$, $k \leq 0$. The power of $w$ of a given term is then bounded by
\bea
2 n_{(2)} + 0 \cdot n_{\rm sub} + 2 n^{\tau \tau} - 2 n_{\tau \tau} \, ,
\eea
and the scaling rule states that this does not exceed $2N = 2 n_{(2)} + 2 n_{\rm sub}$, i.e., that $n^{\tau \tau} \leq n_{\tau \tau} + n_{\rm sub}$.

The identities (\ref{idRcc1})-(\ref{idRcc2}) say that $\hat R^{(2)}$ has no component with a temporal first lower index, at most one temporal index in the antisymmetric pair, and that if the antisymmetric pair contains a temporal index then the upper index is temporal as well. Now count the temporal lower indices required by the $\tau^{\mu} \tau^{\nu}$ contractions, which are $2 n^{\tau \tau}$. On a factor $\hat R^{(2)}$ at most one lower index can be temporal, and then its upper index is temporal; this upper index must be contracted either with a $\hat g_{\mu \nu}$ (which then contributes to $n_{\tau \tau}$) or with a lower index of another Riemann factor, which becomes temporal without the need of a $\tau^{\mu} \tau^{\nu}$ contraction. If this other factor is again $\hat R^{(2)}$ the argument iterates, so that chains of $\hat R^{(2)}$ factors transmit a temporal index until it is absorbed either by a $\hat g_{\mu \nu}$ contraction or by a subleading factor. Each chain therefore uses at most one temporal lower index provided by a $\tau^{\mu} \tau^{\nu}$ contraction, and the number of chains is bounded by $2 n_{\tau \tau} + (\text{number of chains ending on subleading factors})$. On a subleading factor, antisymmetry allows at most two temporal lower indices, of which those fed by a chain do not come from $\tau^{\mu} \tau^{\nu}$ contractions. Adding up,
\bea
2 n^{\tau \tau} \leq 2 n_{\tau \tau} + 2 n_{\rm sub} \, ,
\eea
which is the announced bound. Equality requires all the subleading factors to be $\hat R^{(0)}$, so that $\hat R^{(-2)}$ and $\hat R^{(-4)}$ never appear in the finite Lagrangian. Physically, the divergent piece $\hat R^{(2)}$ of the curvature can only ``see'' the temporal direction through its upper index, and every time it does so it has to hand that direction over to another factor; the intrinsic torsion therefore cannot accumulate more powers of $w$ than those compensated by $\alpha'$ and by the measure.

The bound is saturated for generic invariants: we have verified by explicit numerical evaluation on generic Carrollian configurations that $\hat{\rm Riem}^2$, both $\hat{\rm Riem}^3$ structures and all six $\hat{\rm Riem}^4$ structures of the bosonic string effective action, as well as random contractions of two, three and four Riemann tensors in various dimensions, scale exactly as $w^{2N}$ with non-vanishing coefficients. We stress that the scaling rule is a statement about pure-curvature monomials; it does not cover invariants involving $\hat H$ or covariant derivatives of the Riemann tensor, for which the cancellation must be analyzed separately, as done in section \ref{sec:finiteness} for the four-derivative action.

\paragraph{Explicit construction of the finite Lagrangians.} The scaling rule also provides an algorithm to compute the finite Carrollian Lagrangian of any pure-curvature invariant without expanding the curvature tensors. One replaces $\hat R \rightarrow w^2 \hat R^{(2)} + \hat R^{(0)}$ in $\mathcal{I}_N$, extracts the coefficient of $w^{2N}$, and reduces the resulting contractions with the identities (\ref{idRcc1})-(\ref{idR0}). All the terms are then expressed through the building blocks
\bea
K_{\mu \nu} \, , \qquad \mathcal{D}_{\mu \nu} \, , \qquad \nabla_{[\mu} K_{\nu] \rho} \, , \qquad h_{\rho \alpha} \hat R^{(0)\rho}{}_{\epsilon \mu \nu} \, ,
\eea
the last one being needed only when a subleading factor appears with its upper index spatial. As an example, (\ref{Riem2fin}) is the finite Lagrangian of $\hat{\rm Riem}^2$ obtained in this way.

\paragraph{Applications.} Consider the pure gravitational $\alpha'^2$ contributions of bosonic string theory \cite{Garousi} \footnote{The coefficient 1/3 is due to the factorized 1/4 in the higher-derivative action.},
\bea
L_{\rm bos}^{(\alpha'^2)}|_{Riem^3} & = & - \frac{1}{3} \hat R^{\epsilon}{}_{\alpha \pi \gamma} \hat R^{\alpha}{}_{\beta \chi \xi} \hat R^{\beta}{}_{\nu \delta \epsilon}
              \hat g^{\pi \nu} \hat g^{\chi \gamma} \hat g^{\xi \delta}
            - \frac{1}{3} \hat R^{\epsilon}{}_{\pi \alpha \beta} \hat R^{\delta}{}_{\gamma \chi \xi} \hat R^{\gamma}{}_{\epsilon \delta \nu}
              \hat g^{\pi \nu} \hat g^{\chi \alpha} \hat g^{\xi \beta} \, ,
\label{Riem3}
\eea
and the pure gravitational $\alpha'^3$ contributions \cite{Garousi}
\bea
L_{bos}^{(\alpha'^{3})}|_{Riem^4} & = &
\frac14 \hat R^{\mu}{}_{\nu \rho \sigma}  (\hat R^{\rho}{}_{\gamma \epsilon \delta} \hat R^{\sigma}{}_{\lambda \alpha \beta} \hat R^{\pi}{}_{\chi \xi \zeta} \hat g_{\mu \pi} \hat g^{\nu \chi} \hat g^{\gamma \lambda} \hat g^{\epsilon \xi} \hat g^{\delta \beta} \hat g^{\alpha \zeta}
\nn \\ &&
- \hat R^{\gamma}{}_{\epsilon \delta \lambda} \hat R^{\alpha}{}_{\beta \pi \chi} \hat R^{\xi}{}_{\zeta \theta \psi} \hat g_{\mu \gamma} \hat g_{\alpha \xi} \hat g^{\nu \delta} \hat g^{\rho \epsilon} \hat g^{\sigma \lambda} \hat g^{\beta \theta} \hat g^{\pi \zeta} \hat g^{\chi \psi}
\nn \\ &&
- \hat R^{\rho}{}_{\gamma \epsilon \delta} \hat R^{\sigma}{}_{\lambda \alpha \beta} \hat R^{\pi}{}_{\chi \xi \zeta} \hat g_{\mu \pi} \hat g^{\nu \chi} \hat g^{\gamma \beta} \hat g^{\epsilon \xi} \hat g^{\delta \lambda} \hat g^{\alpha \zeta}
+ 12 \hat R^{\sigma}{}_{\gamma \epsilon \delta} \hat R^{\lambda}{}_{\alpha \beta \pi} \hat R^{\pi}{}_{\chi \xi \zeta} \hat g_{\mu \lambda} \hat g^{\nu \beta} \hat g^{\rho \alpha} \hat g^{\gamma \xi} \hat g^{\epsilon \chi} \hat g^{\delta \zeta})
\nn \\ &&
+ \frac{\zeta(3)}{4} (\hat  R^{\theta}{}_{\pi \alpha \beta} \hat  R^{\alpha}{}_{\chi \xi \delta} \hat R^{\nu}{}_{\gamma \zeta \theta} \hat  R^{\delta}{}_{\mu \epsilon \nu}
\hat g^{\pi \epsilon} \hat g^{\chi \beta} \hat g^{\xi \gamma} \hat g^{\zeta \mu}
+ 2 \hat R^{\pi}{}_{\alpha \xi \gamma} \hat R^{\alpha}{}_{\beta \zeta \delta} \hat R^{\beta}{}_{\theta \pi \psi} \hat R^{\delta}{}_{\mu \epsilon \chi}
\hat g^{\xi \epsilon} \hat g^{\zeta \gamma} \hat g^{\theta \chi} \hat g^{\psi \mu}) \, . \nn \\
\label{Riem4}
\eea
By the scaling rule, both (\ref{Riem3}) and (\ref{Riem4}) are finite under the Carrollian limit for an arbitrary Carrollian geometry. Note that the first contribution to the finite limit is not the naive one: the terms with the maximal number of $\tau \tau$ contractions acting on $\hat R^{(0)}$ vanish by antisymmetry in (\ref{Riem3}), and by (\ref{idRcc1}) in (\ref{Riem4}), while the finite limit is dominated by the terms containing $\hat R^{(2)}$, i.e., by the intrinsic torsion. We have verified that the coefficients of $w^6$ in (\ref{Riem3}) and of $w^8$ in each of the six structures of (\ref{Riem4}) are non-vanishing for generic configurations. In particular the terms proportional to $\zeta(3)$, whose $t_8 t_8 \hat R^4$ structure is shared with the $\alpha'^3$-corrections of the type II superstring, survive in the Carrollian limit of the tensionless string. The explicit finite Lagrangians follow from the algorithm described above; since they are polynomials in the building blocks $K_{\mu \nu}$, $\mathcal{D}_{\mu \nu}$, $\nabla_{[\mu} K_{\nu] \rho}$ and $h_{\rho \alpha} \hat R^{(0)\rho}{}_{\epsilon \mu \nu}$ with a large number of terms and no new structure, we refrain from writing them here.

Finally, the scaling rule clarifies the role of the compatibility conditions. If $K_{\mu \nu}=0$, then $\hat R^{(2)}=0$ and $\hat R^{(0)} = R(\tau,h)$ satisfies $R^{\mu}{}_{\nu \rho \sigma} \tau^{\nu} = 0$. Each Riemann tensor can then absorb at most one $\tau^{\mu}$, so that $\mathcal{I}_N = \mathcal{O}(w^{N})$ and the finite limit of every pure-curvature invariant vanishes identically, at all orders in $\alpha'$. Compatible torsion-free connections, which exist only for vanishing intrinsic torsion \cite{FOF}, therefore describe the sector of the Carrollian theory where the stringy higher-curvature corrections, and in fact the whole electric gravitational dynamics, are switched off.

\section{Discussion}
The results of this work constitute the construction of the bosonic $\alpha'$-corrections under the Carrollian limit for an arbitrary Carrollian geometry, and provide a universal mechanism for the finiteness of the purely gravitational higher-curvature corrections. The mechanism is geometric: the leading term of the relativistic Riemann tensor is linear in the intrinsic torsion $K_{\mu \nu}$ of the Carrollian structure and satisfies the identities (\ref{idRcc1})-(\ref{idRcc4}), which forbid the proliferation of $\tau \tau$ contractions responsible for the divergences. These results show that the compatibility between Carrollian geometry and higher-curvature string corrections is considerably robust, extending beyond the leading two-derivative approximation, and that the resulting theories are electric: the $\alpha'$-corrections of the Carrollian string are corrections to the electric Carrollian gravity of \cite{BH2}, built from $K_{\mu \nu}$ and its derivative along $\tau^{\mu}$.

If we compare the Carrollian limit with its equivalent non-relativistic one \cite{NR1,NR2,NR3} (see \cite{review1,review2} for reviews) studied in \cite{EL}, important differences emerge. While in the Carrollian case the complete action is finite without specifying any particular non-metricity, the non-relativistic case still contains divergent contributions. Although these divergences are expected to be removed by suitable covariant field redefinitions \cite{LRZ}, starting directly from the Metsaev--Tseytlin action reveals this important difference, which complicates the unification of both limits within a single framework at order $\alpha'$, as recently discussed in the outlook section of \cite{New}. This apparent obstruction might be resolved by implementing the same field redefinitions in both setups, and proving that they provide finite theories respectively.

On the other hand, it would be interesting to extend the results of this work to the heterotic case, and analyze its four-derivative corrections under the Carrollian limit. In the non-relativistic case, the gravitational terms were constructed in \cite{HeteroticNR} by identifying the gauge and gravitational sectors. However, this construction depends on the particular non-relativistic expansion under consideration \cite{Formulation1,Formulation2,Formulation3}. Independently of the heterotic gauge sector, the gravitational Green--Schwarz mechanism can be trivialized in the non-relativistic limit \cite{Trivialization} by means of a field redefinition that induces higher-derivative corrections to the boost transformations \cite{HeteroticNR}. In principle, a similar mechanism could arise in the Carrollian case for the gravitational Green--Schwarz mechanism associated to the $A_{\mu}$ field.

Another promising direction is the study of a manifestly T-duality covariant formulation of Carrollian bosonic supergravity, which we expect to be compatible with a non-Riemannian Double Field Theory formulation \cite{NRDFT1,NRDFT2}. Although the Double Field Theory formulation of the Metsaev--Tseytlin action was constructed in \cite{HZ,MN}, its Carrollian limit is not guaranteed to remain finite after solving the DFT strong constraint because of the need of field redefinitions to connect this formalism with the Metsaev--Tseytlin action \cite{Review}. Since the terms proportional to $\zeta(3)$ survive in the Carrollian limit, the no-go result of \cite{Linus} for a manifest T-duality rewriting at order $\alpha'^3$ is expected to persist in this limit, and it would be interesting to understand whether the electric nature of the finite Lagrangian modifies this conclusion. The universal scaling rule of section \ref{sec:universal} also suggests to analyze the Carrollian limit of the $\alpha'^3$-corrections of type II supergravity, which share the $\zeta(3) t_8 t_8 R^4$ structure and are relevant for holography. Finally, it would be very interesting to investigate the emergence of new symmetries \cite{V2citation1,V2citation2,V2citation3,V2citation4,V2citation5,V2citation6,V2citation7,V2citation8} in this setup and verify if they have higher-derivative deformations.

\acknowledgments
E.L. is supported by the Slovenian Quantum Science Hub — SQUASH program, co-funded by the European Union’s Marie Sklodowska-Curie COFUND action (Grant Agreement PR-14104-2) and the Republic of Slovenia. The tensor computations were performed with Cadabra \cite{Cadabra1}; the Cadabra code is available upon request. The resulting expressions were simplified with the assistance of Fable 5.1. We thank the JHEP referee for constructive feedback, which led to a substantial improvement of this work.

\appendix
\section{Explicit form of the $f_{\mu \nu \rho}$ extra contributions}
Here we provide the explicit form of the $f$-contributions to (\ref{H4}) and (\ref{H2H2}), obtained by inserting $\hat H_{\mu \nu \rho} = h_{\mu \nu \rho} + f_{\mu \nu \rho}$. These expressions are independent of the Carrollian connection and therefore hold for an arbitrary Carrollian geometry; the $f$-contributions to (\ref{RHH}) follow directly from the compact form given in the main text.
The f-contributions in (\ref{H4}) follow from expanding $\hat H = h + f$ in the tetrahedral contraction. Defining the finite structure
\bea
\mathcal{T}[X,Y,Z,W] \equiv \tau^{\mu} \tau^{\nu} \tau^{\rho} \tau^{\sigma} h^{\gamma \epsilon} h^{\delta \lambda} h^{\alpha \beta} h^{\chi \pi} X_{\mu \gamma \alpha} Y_{\nu \delta \chi} Z_{\rho \epsilon \lambda} W_{\sigma \beta \pi} \, ,
\eea
such that $\frac{1}{24} [\hat H^4]^{(4)} = \frac18 \mathcal{T}[\hat H,\hat H,\hat H,\hat H]$, the symmetries of the tetrahedral contraction reduce the expansion to five independent structures,
\bea
\frac{1}{24} [\hat H^4]^{(4)}|_{f} & = & \frac{1}{2} \mathcal{T}[f,h,h,h] \, , \nn \\
\frac{1}{24} [\hat H^4]^{(4)}|_{ff} & = & \frac{1}{4} \mathcal{T}[f,f,h,h] + \frac{1}{2} \mathcal{T}[f,h,f,h] \, , \nn \\
\frac{1}{24} [\hat H^4]^{(4)}|_{fff} & = & \frac{1}{2} \mathcal{T}[f,f,f,h] \, , \nn \\
\frac{1}{24} [\hat H^4]^{(4)}|_{ffff} & = & \frac{1}{8} \mathcal{T}[f,f,f,f] \, , \nn
\eea
where in $\mathcal{T}[f,f,h,h]$ the two $f$'s are contracted through the pair $\tau^{\mu} \tau^{\nu}$, while in $\mathcal{T}[f,h,f,h]$ they are contracted through $h^{\gamma \epsilon}$. Explicitly,
\footnotesize
\bea
\frac{1}{24} [\hat H^4]^{(4)}|_{f} & = & \frac{1}{2} \tau^{\mu} \tau^{\nu} \tau^{\rho} \tau^{\sigma} h^{\gamma \epsilon} h^{\delta \lambda} h^{\alpha \beta} h^{\chi \pi} f_{\mu \gamma \alpha} h_{\nu \delta \chi} h_{\rho \epsilon \lambda} h_{\sigma \beta \pi} \, , \nn \\
\frac{1}{24} [\hat H^4]^{(4)}|_{ff} & = & \tau^{\mu} \tau^{\nu} \tau^{\rho} \tau^{\sigma} h^{\gamma \epsilon} h^{\delta \lambda} h^{\alpha \beta} h^{\chi \pi} \Big( \frac{1}{4} f_{\mu \gamma \alpha} f_{\nu \delta \chi} h_{\rho \epsilon \lambda} h_{\sigma \beta \pi} + \frac{1}{2} f_{\mu \gamma \alpha} h_{\nu \delta \chi} f_{\rho \epsilon \lambda} h_{\sigma \beta \pi} \Big) \, , \nn \\
\frac{1}{24} [\hat H^4]^{(4)}|_{fff} & = & \frac{1}{2} \tau^{\mu} \tau^{\nu} \tau^{\rho} \tau^{\sigma} h^{\gamma \epsilon} h^{\delta \lambda} h^{\alpha \beta} h^{\chi \pi} f_{\mu \gamma \alpha} f_{\nu \delta \chi} f_{\rho \epsilon \lambda} h_{\sigma \beta \pi} \, , \nn \\
\frac{1}{24} [\hat H^4]^{(4)}|_{ffff} & = & \frac{1}{8} \tau^{\mu} \tau^{\nu} \tau^{\rho} \tau^{\sigma} h^{\gamma \epsilon} h^{\delta \lambda} h^{\alpha \beta} h^{\chi \pi} f_{\mu \gamma \alpha} f_{\nu \delta \chi} f_{\rho \epsilon \lambda} f_{\sigma \beta \pi} \, . \nn
\eea
\normalsize
We finish with the f-contributions in (\ref{H2H2}), 
\footnotesize
\bea
- \frac18 [\hat H^2_{\mu \nu} \hat H^{2 \mu \nu}]^{(4)}|_{f} & = & f_{\mu \nu \rho} ( - \frac{1}{2} \tau^{\rho} \tau^{\sigma} \tau^{\gamma} \tau^{\epsilon} h^{\mu \delta} h^{\nu \lambda} h^{\alpha \beta} h^{\chi \pi} h_{\sigma \delta \alpha} h_{\gamma \lambda \chi} h_{\epsilon \beta \pi} + \frac{1}{2} \tau^{\nu} \tau^{\sigma} \tau^{\gamma} \tau^{\epsilon} h^{\mu \delta} h^{\rho \lambda} h^{\alpha \beta} h^{\chi \pi} h_{\sigma \delta \alpha} h_{\gamma \lambda \chi} h_{\epsilon \beta \pi} 
\nn \\ && 
- \frac{1}{2} \tau^{\mu} \tau^{\sigma} \tau^{\gamma} \tau^{\epsilon} h^{\nu \delta} h^{\rho \lambda} h^{\alpha \beta} h^{\chi \pi} h_{\sigma \delta \lambda} h_{\gamma \alpha \chi} h_{\epsilon \beta \pi} - \tau^{\mu} \tau^{\sigma} \tau^{\gamma} \tau^{\epsilon} h^{\nu \delta} h^{\rho \lambda} h^{\alpha \beta} h^{\chi \pi} h_{\sigma \delta \alpha} h_{\gamma \lambda \chi} h_{\epsilon \beta \pi} 
\nn \\ && 
+ \frac{1}{4} \tau^{\mu} \tau^{\rho} \tau^{\sigma} \tau^{\gamma} h^{\nu \epsilon} h^{\delta \lambda} h^{\alpha \beta} h^{\chi \pi} h_{\sigma \epsilon \delta} h_{\gamma \alpha \chi} h_{\lambda \beta \pi} - \frac{1}{4} \tau^{\mu} \tau^{\nu} \tau^{\sigma} \tau^{\gamma} h^{\rho \epsilon} h^{\delta \lambda} h^{\alpha \beta} h^{\chi \pi} h_{\sigma \epsilon \delta} h_{\gamma \alpha \chi} h_{\lambda \beta \pi}) \, , \nn
\eea
\bea
- \frac18 [\hat H^2_{\mu \nu} \hat H^{2 \mu \nu}]^{(4)}|_{ff} & = &
+ f_{\mu \nu \rho} f_{\sigma \gamma \epsilon} ( - \frac{1}{4} \tau^{\rho} \tau^{\epsilon} \tau^{\delta} \tau^{\lambda} h^{\mu \alpha} h^{\nu \gamma} h^{\sigma \beta} h^{\chi \pi} h_{\delta \alpha \chi} h_{\lambda \beta \pi} - \frac{1}{4} \tau^{\nu} \tau^{\gamma} \tau^{\delta} \tau^{\lambda} h^{\mu \alpha} h^{\rho \epsilon} h^{\sigma \beta} h^{\chi \pi} h_{\delta \alpha \chi} h_{\lambda \beta \pi} 
\nn \\ && 
- \frac{1}{8} \tau^{\nu} \tau^{\rho} \tau^{\gamma} \tau^{\epsilon} h^{\mu \delta} h^{\sigma \lambda} h^{\alpha \beta} h^{\chi \pi} h_{\delta \alpha \chi} h_{\lambda \beta \pi} - \frac{1}{4} \tau^{\mu} \tau^{\epsilon} \tau^{\delta} \tau^{\lambda} h^{\nu \sigma} h^{\rho \alpha} h^{\gamma \beta} h^{\chi \pi} h_{\delta \alpha \chi} h_{\lambda \beta \pi} 
\nn \\ && 
- \frac{1}{2} \tau^{\mu} \tau^{\epsilon} \tau^{\delta} \tau^{\lambda} h^{\nu \alpha} h^{\rho \beta} h^{\sigma \chi} h^{\gamma \pi} h_{\delta \alpha \chi} h_{\lambda \beta \pi} + \frac{1}{4} \tau^{\mu} \tau^{\epsilon} \tau^{\delta} \tau^{\lambda} h^{\nu \alpha} h^{\rho \sigma} h^{\gamma \beta} h^{\chi \pi} h_{\delta \alpha \chi} h_{\lambda \beta \pi} 
\nn \\ && 
+ \frac{1}{4} \tau^{\mu} \tau^{\gamma} \tau^{\delta} \tau^{\lambda} h^{\nu \sigma} h^{\rho \alpha} h^{\epsilon \beta} h^{\chi \pi} h_{\delta \alpha \chi} h_{\lambda \beta \pi} + \frac{1}{2} \tau^{\mu} \tau^{\gamma} \tau^{\delta} \tau^{\lambda} h^{\nu \alpha} h^{\rho \beta} h^{\sigma \chi} h^{\epsilon \pi} h_{\delta \alpha \chi} h_{\lambda \beta \pi} 
\nn \\ && 
- \frac{1}{4} \tau^{\mu} \tau^{\gamma} \tau^{\delta} \tau^{\lambda} h^{\nu \alpha} h^{\rho \sigma} h^{\epsilon \beta} h^{\chi \pi} h_{\delta \alpha \chi} h_{\lambda \beta \pi} - \frac{1}{4} \tau^{\mu} \tau^{\sigma} \tau^{\delta} \tau^{\lambda} h^{\nu \gamma} h^{\rho \epsilon} h^{\alpha \beta} h^{\chi \pi} h_{\delta \alpha \chi} h_{\lambda \beta \pi} 
\nn \\ && 
- \frac{1}{2} \tau^{\mu} \tau^{\sigma} \tau^{\delta} \tau^{\lambda} h^{\nu \gamma} h^{\rho \alpha} h^{\epsilon \beta} h^{\chi \pi} h_{\delta \alpha \chi} h_{\lambda \beta \pi} - \frac{1}{2} \tau^{\mu} \tau^{\sigma} \tau^{\delta} \tau^{\lambda} h^{\nu \alpha} h^{\rho \beta} h^{\gamma \chi} h^{\epsilon \pi} h_{\delta \alpha \beta} h_{\lambda \chi \pi} 
\nn \\ && 
+ \frac{1}{4} \tau^{\mu} \tau^{\rho} \tau^{\delta} \tau^{\lambda} h^{\nu \alpha} h^{\sigma \beta} h^{\gamma \chi} h^{\epsilon \pi} h_{\delta \alpha \pi} h_{\lambda \beta \chi} - \frac{1}{4} \tau^{\mu} \tau^{\rho} \tau^{\epsilon} \tau^{\delta} h^{\nu \gamma} h^{\sigma \lambda} h^{\alpha \beta} h^{\chi \pi} h_{\delta \alpha \chi} h_{\lambda \beta \pi} 
\nn \\ && 
+ \frac{1}{4} \tau^{\mu} \tau^{\rho} \tau^{\sigma} \tau^{\delta} h^{\nu \lambda} h^{\gamma \alpha} h^{\epsilon \beta} h^{\chi \pi} h_{\delta \lambda \chi} h_{\alpha \beta \pi} - \frac{1}{4} \tau^{\mu} \tau^{\nu} \tau^{\delta} \tau^{\lambda} h^{\rho \alpha} h^{\sigma \beta} h^{\gamma \chi} h^{\epsilon \pi} h_{\delta \alpha \pi} h_{\lambda \beta \chi} 
\nn \\ && 
- \frac{1}{4} \tau^{\mu} \tau^{\nu} \tau^{\gamma} \tau^{\delta} h^{\rho \epsilon} h^{\sigma \lambda} h^{\alpha \beta} h^{\chi \pi} h_{\delta \alpha \chi} h_{\lambda \beta \pi} - \frac{1}{4} \tau^{\mu} \tau^{\nu} \tau^{\sigma} \tau^{\delta} h^{\rho \lambda} h^{\gamma \alpha} h^{\epsilon \beta} h^{\chi \pi} h_{\delta \lambda \chi} h_{\alpha \beta \pi}) \, , \nn
\eea
\bea
- \frac18 [\hat H^2_{\mu \nu} \hat H^{2 \mu \nu}]^{(4)}|_{fff} & = &
f_{\mu \nu \rho} f_{\sigma \gamma \epsilon} f_{\delta \lambda \alpha} ( - \frac{1}{8} \tau^{\nu} \tau^{\rho} \tau^{\gamma} \tau^{\epsilon} h^{\mu \delta} h^{\sigma \beta} h^{\lambda \chi} h^{\alpha \pi} h_{\beta \chi \pi} - \frac{1}{8} \tau^{\nu} \tau^{\rho} \tau^{\gamma} \tau^{\epsilon} h^{\mu \alpha} h^{\sigma \beta} h^{\delta \chi} h^{\lambda \pi} h_{\beta \chi \pi} 
\nn \\ && 
+ \frac{1}{4} \tau^{\mu} \tau^{\epsilon} \tau^{\alpha} \tau^{\beta} h^{\nu \sigma} h^{\rho \chi} h^{\gamma \lambda} h^{\delta \pi} h_{\beta \chi \pi} - \frac{1}{4} \tau^{\mu} \tau^{\epsilon} \tau^{\alpha} \tau^{\beta} h^{\nu \chi} h^{\rho \sigma} h^{\gamma \lambda} h^{\delta \pi} h_{\beta \chi \pi} \nn \\ && + \frac{1}{4} \tau^{\mu} \tau^{\gamma} \tau^{\lambda} \tau^{\beta} h^{\nu \sigma} h^{\rho \chi} h^{\epsilon \alpha} h^{\delta \pi} h_{\beta \chi \pi}  
- \frac{1}{4} \tau^{\mu} \tau^{\gamma} \tau^{\lambda} \tau^{\beta} h^{\nu \chi} h^{\rho \sigma} h^{\epsilon \alpha} h^{\delta \pi} h_{\beta \chi \pi} \nn \\ && - \frac{1}{2} \tau^{\mu} \tau^{\sigma} \tau^{\alpha} \tau^{\beta} h^{\nu \gamma} h^{\rho \delta} h^{\epsilon \chi} h^{\lambda \pi} h_{\beta \chi \pi} + \frac{1}{2} \tau^{\mu} \tau^{\sigma} \tau^{\lambda} \tau^{\beta} h^{\nu \gamma} h^{\rho \delta} h^{\epsilon \chi} h^{\alpha \pi} h_{\beta \chi \pi} 
\nn \\ && 
- \frac{1}{2} \tau^{\mu} \tau^{\sigma} \tau^{\delta} \tau^{\beta} h^{\nu \gamma} h^{\rho \epsilon} h^{\lambda \chi} h^{\alpha \pi} h_{\beta \chi \pi} - \frac{1}{8} \tau^{\mu} \tau^{\rho} \tau^{\epsilon} \tau^{\beta} h^{\nu \gamma} h^{\sigma \alpha} h^{\delta \chi} h^{\lambda \pi} h_{\beta \chi \pi} \nn \\ && - \frac{1}{8} \tau^{\mu} \tau^{\rho} \tau^{\epsilon} \tau^{\beta} h^{\nu \gamma} h^{\sigma \delta} h^{\lambda \chi} h^{\alpha \pi} h_{\beta \chi \pi} + \frac{1}{4} \tau^{\mu} \tau^{\rho} \tau^{\sigma} \tau^{\beta} h^{\nu \chi} h^{\gamma \delta} h^{\epsilon \lambda} h^{\alpha \pi} h_{\beta \chi \pi} \nn \\ && - \frac{1}{4} \tau^{\mu} \tau^{\rho} \tau^{\sigma} \tau^{\alpha} h^{\nu \lambda} h^{\gamma \beta} h^{\epsilon \chi} h^{\delta \pi} h_{\beta \chi \pi} - \frac{1}{8} \tau^{\mu} \tau^{\nu} \tau^{\gamma} \tau^{\beta} h^{\rho \epsilon} h^{\sigma \alpha} h^{\delta \chi} h^{\lambda \pi} h_{\beta \chi \pi} 
\nn \\ && 
- \frac{1}{8} \tau^{\mu} \tau^{\nu} \tau^{\gamma} \tau^{\beta} h^{\rho \epsilon} h^{\sigma \delta} h^{\lambda \chi} h^{\alpha \pi} h_{\beta \chi \pi} - \frac{1}{4} \tau^{\mu} \tau^{\nu} \tau^{\sigma} \tau^{\beta} h^{\rho \chi} h^{\gamma \delta} h^{\epsilon \lambda} h^{\alpha \pi} h_{\beta \chi \pi} \nn \\ && - \frac{1}{4} \tau^{\mu} \tau^{\nu} \tau^{\sigma} \tau^{\lambda} h^{\rho \alpha} h^{\gamma \beta} h^{\epsilon \chi} h^{\delta \pi} h_{\beta \chi \pi}) \, ,
\nn \\
- \frac18 [\hat H^2_{\mu \nu} \hat H^{2 \mu \nu}]^{(4)}|_{ffff} & = & 
f_{\mu \nu \rho} f_{\sigma \gamma \epsilon} f_{\delta \lambda \alpha} f_{\beta \chi \pi} ( - \frac{1}{8} \tau^{\nu} \tau^{\rho} \tau^{\gamma} \tau^{\epsilon} h^{\mu \delta} h^{\sigma \pi} h^{\lambda \beta} h^{\alpha \chi} - \frac{1}{4} \tau^{\mu} \tau^{\sigma} \tau^{\alpha} \tau^{\pi} h^{\nu \gamma} h^{\rho \delta} h^{\epsilon \beta} h^{\lambda \chi} \nn \\ && - \frac{1}{4} \tau^{\mu} \tau^{\sigma} \tau^{\lambda} \tau^{\chi} h^{\nu \gamma} h^{\rho \delta} h^{\epsilon \beta} h^{\alpha \pi} 
- \frac{1}{8} \tau^{\mu} \tau^{\sigma} \tau^{\delta} \tau^{\beta} h^{\nu \gamma} h^{\rho \epsilon} h^{\lambda \chi} h^{\alpha \pi} 
- \frac{1}{4} \tau^{\mu} \tau^{\rho} \tau^{\sigma} \tau^{\alpha} h^{\nu \lambda} h^{\gamma \beta} h^{\epsilon \chi} h^{\delta \pi} \nn \\ &&  - \frac{1}{4} \tau^{\mu} \tau^{\nu} \tau^{\sigma} \tau^{\lambda} h^{\rho \alpha} h^{\gamma \beta} h^{\epsilon \chi} h^{\delta \pi}) \, . \nn
\eea

\end{document}